Title:

# Deep Learning Angiography (DLA): Three-dimensional C-arm Cone Beam CT Angiography Using Deep Learning

Juan C. Montoya[1]
Yinsheng Li[1]
Charles Strother, MD[2]
Guang-Hong Chen, PhD [*1,2]

[1]Department of Medical Physics, University of Wisconsin School of Medicine and Public Health, Madison, WI

[2]Department of Radiology, University of Wisconsin School of Medicine and Public Health, Madison, WI

**\*Corresponding Author**
Guang-Hong Chen, PhD
L1167, WIMR
1111 Highland Ave, 1167
Madison, WI 53705, USA
Phone:608-263-0089
Fax:608-265-9840
Email: gchen7@wisc.edu

**Financial Disclosure**: Dr. Guang-Hong Chen received funding support from Siemens Healthineers for unrelated research project. Dr. Charles Strother received funding support from Siemens Healthineers for unrelated research project. The work is partially supported by a National Institutes of Health grant U01EB021183.

# Abstract

**Background and Purpose**

Our purpose was to develop a deep learning angiography (DLA) method to generate 3D cerebral angiograms from a single contrast-enhanced acquisition.

**Material and Methods**

Under an approved IRB protocol 105 3D-DSA exams were randomly selected from an internal database. All were acquired using a clinical system (Axiom Artis zee; Siemens Healthineers) in conjunction with a standard injection protocol. More than 150 million labeled voxels from 35 subjects were used for training. A deep convolutional neural network was trained to classify each image voxel into three tissue types (vasculature, bone and soft tissue). The trained DLA model was then applied for tissue classification in a validation cohort of 8 subjects and a final testing cohort consisting of the remaining 62 subjects. The final vasculature tissue class was used to generate the 3D-DLA images. To quantify the generalization error of the trained model, accuracy, sensitivity, precision and F1-scores were calculated for vasculature classification in relevant anatomy. The 3D-DLA and clinical 3D-DSA images were subject to a qualitative assessment for the presence of inter-sweep motion artifacts.

**Results**

Vasculature classification accuracy and 95% CI in the testing dataset was 98.7% ([98.3, 99.1] %). No residual signal from osseous structures was observed for all 3D-DLA testing cases except for small regions in the otic capsule and nasal cavity compared to 37% (23/62) of the 3D-DSAs.

**Conclusion**

DLA accurately recreated the vascular anatomy of the 3D-DSA reconstructions without mask. DLA reduced mis-registration artifacts induced by inter-sweep motion. DLA reduces radiation exposure required to obtain clinically useful 3D-DSA.

**Abbreviations**

DLA – Deep learning angiography

CNN – Convolutional Neural Network

**Introduction**

Cerebrovascular diseases are common causes of morbidity and mortality in the adult population worldwide[1-3]. Most cerebrovascular diseases are found during routine brain imaging with CT or MRI, however 2D-DSA remains the gold standard for their accurate angiographic evaluation and characterization, in particular for arteriovenous malformations [4], cerebral aneurysms [5,6] and dural arteriovenous fistulas [7]. Additional 3D rotational angiography (3D-DSA) is used to improve the visualization and spatial understanding of vascular structures during the diagnostic work-up of these conditions. Currently, with many angiographic systems, obtaining a 3D-DSA still requires two rotational acquisitions; one without injection of contrast (mask run) and one during injection of contrast (fill run). These two datasets are used to compute log-subtracted projections which are then used to reconstruct a subtracted 3D-DSA volume [8,9].

Machine learning is a discipline within computer science, closely related to statistics and mathematical optimization, that aims to learn patterns directly from a large set of examples that demonstrate a desired outcome or behavior without the need of explicit instructions [10]. In the context of medical imaging, machine learning methods have been investigated since early 1990s, initially for computed aided detection and diagnosis in mammography and pulmonary embolism [11-14], however recent advances in deep learning [15] (i.e. a specific machine learning technique) have demonstrated unprecedented performance in many applications, including detections of diabetic retinopathy [16], breast cancer [17,18], quantitative analysis of brain tumors in MRI [19,20], computed-aided detection of cerebral aneurysms in MR angiography[21] and computer aided detection and classification of thoracic diseases [22, 23].

With recent advances in deep learning and the universal approximation properties of feedforward neural networks [24, 25], it is hypothesized that a deep neural network is capable of computing cerebral angiograms with only the vascular information contained in the fill scan of a 3D-DSA exam acquired with a C-arm cone beam CT system. If possible, potential benefits of eliminating the mask scan include 1) reduction of inter-sweep patient motion artifacts caused by the mis-registration of the mask and fill scans and 2) radiation dose reduction by at least a factor of 2.

The purpose of this work was to develop and test the capability of a deep learning angiography (DLA) method, based on convolutional neural networks (CNN), to generate subtracted 3D cerebral angiograms from a single contrast-enhanced exam without the need for a mask acquisition.

**Methods**

In the following sections, the patient inclusion criteria and image acquisition protocols are first presented, followed by a description of the datasets and methods used to train the DLA model. Finally, the image analysis and statistical analysis are described. The overall study schema is shown in Figure 1.

**Patient Cohort**

All studies were HIPAA compliant and done under an Institutionally Review Board approved protocol. Clinically indicated rotational angiography exams for the assessment of cerebrovascular abnormalities of 105 patients, scanned from August 2014 through April 2016 were retrospectively collected. Cases were selected in a random fashion to reduce the potential bias in patient selection. It was thought that the randomized selection over this period would result in a data set that was representative of the varieties of conditions that are referred for angiographic studies.

**Imaging Acquisition and Reconstruction**

All subjects were imaged with a standard 3D-DSA data acquisition protocol using a C-arm cone beam CT system (Axiom Artis zee; Siemens Healthineers). The protocol consists of the acquisition of two cone beam CT acquistions(i.e. mask and fill acquisitions) with 172 or 304 projection images for a 6 or 13 second rotation time respectively. Angular coverage for all data acquisition is 260º, with a tube potential of 70 kVp, detector dose per projection image equal to 0.36 µGy per frame and angular increments of 1.52º or 0.85 º per frame. Iodinated contrast medium was injected in the proximal internal carotid artery or vertebral artery just after the initiation of the fill acquisition. For each subject, "native fill" and subtracted 3D volumes were reconstructed using vendor's proprietary software (InSpace Reconstruction, syngo Workplace; Siemens Healthineers). All reconstructions were performed using the standard filtered backprojection with edge enhancement, normal image characteristic, full field of view (238x238 mm$^2$) with 512x512 image matrix, and 0.46 mm image thickness/increment for a 0.46 mm isotropic voxel size. The effective dose for the acquisition protocols used in this study is 1.1 mSv for the 6-second rotation acquisition and 1.8 mSv for 13-second rotation acquisition which is similar to the dose level reported by others[26, 27].

**Training Dataset**

A training dataset consisting of 13790 axial images from 35 patients with over 150 million labeled voxels was generated using the information from both the cone beam CT image of the fill scan and the subtracted images from the subtracted cone beam projection data. For each patient in the training dataset, vasculature extraction was performed by a manual thresholding of the subtracted images. The selection of the threshold was based on the subjective assessment of complete vasculature segmentation while excluding image artifacts and background noise with threshold values typically in the range of 500-700 HU. Large vessels, specifically the internal carotid artery (ICA), middle cerebral artery (MCA), anterior cerebral artery (ACA), distal branches of the MCA and ACA, vertebral artery (VA) and posterior cerebral artery (PCA) were isolated through three-dimensional connected component analysis [28]. Small regions not connected to a large vessel were assumed to be image artifacts and were excluded from the final vasculature volume. After the previous steps, in the event of remaining inter-sweep patient motion and streak artifacts, the vasculature volume was subject to manual artifact removal.

The extraction of bone tissue was performed by subtracting the vasculature volume from the contrast-enhanced images (i.e. fill scan) and performing manual thresholding and connectivity analysis (similar to that of vasculature extraction) in the resulting images. Only connected regions including the skull and mandible were considered to be bone. Remaining streaking artifacts and metal implants in the bone volume were manually removed. Finally, the soft tissue class was extracted by thresholding the fill images with thresholds of [-400,500] HU and applying a morphologic erosion.

The procedure described above generates approximately 0.28 million, 8 million, and 15 million voxels of vasculature, bone, and soft tissue, respectively for each patient. In order to mitigate the class imbalance (i.e. different number of labeled voxels per tissue class) and reduce redundant training data by similarity of adjacent voxels, only 4.3 million labeled voxels per patient were included for training, consisting of all vasculature voxels and equal number of randomly extracted bone and soft tissue voxels (i.e. random undersampling) [29].

**Validation and Testing Datasets**

A validation dataset and a testing dataset were created using the remaining image volumes from 70 subjects divided into 8 exams for the validation dataset and 62 exams for the testing dataset. These datasets were created with the same procedure used to generate the training dataset, however the tissue labels were constrained to a region only containing the following anatomy: ICA, MCA, ACA, distal branches of the MCA and ACA, VA, PCA, the base and anterior aspect of the skull, temporal bone, otic capsule and surrounding soft tissue as opposed to the entire head in the training dataset. Each exam in the validation and testing dataset had approximately the same number of labeled voxels for each tissue class.

**Neural Network Architecture and Implementation**

A 30-layer convolutional neural network (CNN) [30] with a ResNet architecture [31, 32] as shown in Figure 2 was used. All convolutional layers except the input layer, use 3x3 filters with rectified linear units for activation function. The input of the network is a 41x41x5 volumetric image patch extracted from the contrast-enhanced image volume; the network output consists of a 3-way fully connected layer with softmax activation. Training and inference is performed in a voxel-wise basis where the input volumetric image patch is labeled with the tissue class of its central voxel. The DLA model was implemented using TensorFlow (Google Inc, Mountain View, CA). Network parameters were initialized using the variance scaling method [33] and trained from scratch using synchronous stochastic gradient descent method with a batch size of 512 volumetric image patches using 2 GTX 1080 Ti (NVIDIA, Santa Clara, CA) GPUs (256 image patches per GPU). The time required to process one case in this study varies from 1 to 3 minutes, depending on the size of the image volume.

To account for class imbalance, each tissue class had equal probability of being included in a single batch (i.e. data re-sampling) [29]. The learning rate was initially set to be $1 \times 10^{-3}$ with momentum of 0.9. The learning rate was reduced to $1 \times 10^{-4}$ and $1 \times 10^{-5}$ after 1 and 1.5 epochs respectively. The validation dataset was used only to monitor the convergence and generalization error during model training. Early stopping was used when the validation error reached a plateau at $2 \times 10^5$ iterations.

**Statistical Analysis**

The trained DLA model was applied for the task of tissue classification in the, validation and testing cohorts consisting of image volumes from 8 and 62 subjects respectively. The final vasculature tissue class was used to generate the 3D-DLA images. To quantify the generalization error of the trained model, vasculature classification was evaluated for each labeled voxel in the reference standard for the validation and testing datasets. Two-by-two tables were generated for each patient and accuracy, sensitivity (also known as recall), positive predictive value (also known as precision) and dice similarity coefficients (also known as F1-score) were calculated. The 95% CIs for each performance metric were also reported. Finally, the clinical 3D-DSA and the 3D-DLA images were subject to a qualitative assessment for the presence of inter-sweep motion artifacts and results were expressed as frequencies and percentages.

**Results**

Contrast-enhanced image volumes from 105 subjects (age 53.3 ± 13.5 years) who underwent clinically indicated rotational angiography exams for the assessment of cerebrovascular abnormalities were used in the study. Contrast medium was injected via the proximal ICA in 89 patients (85%) and injected via the VA in 16 patients (15%). Average and 95% CIs for vasculature classification accuracy, sensitivity, positive predictive value and Dice similarity coefficient in the testing dataset were 98.7% ([98.3, 99.1] %), 97.6% ([96.5, 98.6] %), 98.5% ([97.6, 99.3] %), 98.0% ([97.4, 98.7] %) respectively. Table 1 summarizes the performance metrics for vascular classification in the training, validation and testing datasets.

No residual signal from osseous structures was observed for all testing cases generated using 3D-DLA except for small regions in the otic capsule and nasal cavity compared to 37% (23/62) of the 3D-DSA cases that presented residual bone artifacts. Figure 3 shows a comparison of MIP images derived from 3D-DSA and the 3D-DLA datasets of a patient evaluated for posterior circulation. One can see how residual bone artifacts induced by inter-sweep patient motion are greatly reduced in 3D-DLA, improving the conspicuity of small vessels. Similarly, Figure 4 shows lateral and oblique MIP images derived from 3D-DSA and the 3D-DLA datasets of a patient evaluated for anterior circulation. Results show reduced residual bone artifacts for 3D-DLA images, in particular for the anterior aspect of the skull and the temporal bone. Figure 5 shows a comparison of volume rendering images for both the clinical 3D-DSA and the 3D-DLA of a patient with a small aneurysm in the anterior communicating artery and a large aneurysm in the MCA bifurcation.

**Discussion**

In this work, a deep CNN was used to learn generic opacified vasculature from contrast enhanced C-arm cone beam CT datasets in order to generate a 3D cerebral angiogram, without explicit definition of cerebrovascular diseases or specific vascular anatomy. The datasets used for model training, validation and testing, were created by applying simple image processing techniques with minimum manual editing to a total of 82740 subtracted and contrast-enhanced cone beam CT images from 105 subjects. The proposed DLA method is used to improve image quality by reducing image artifacts caused by mis-registration of mask and fill scans in 3D-DSA in addition to enabling potential radiation dose reduction.

Many angiographic systems require two rotational acquisitions (mask and fill) for reconstruction of a subtracted 3D-DSA. Others, through the use of vascular segmentation and thresholding algorithms allow a 3D vessel reconstruction to be done without availability of a mask. Those that require two rotations are susceptible to artifacts caused by potential miss-registrations of the mask and fill projections. Those which require use of segmentation and thresholding algorithms may be subject to errors related to too little contrast intensity and/or improper segmentation. Together, these techniques remain the standard of care for the diagnosis and treatment planning of cerebrovascular diseases. Mis-registration artifacts arise in conventional 3D-DSA imaging primarily due to 1) small variations in the angular range differences occurring from one rotational acquisition to another and 2) potential patient motion in both mask and fill runs. The mask-free DLA method, by eliminating the need for one of the rotational acquisitions, in theory, would reduce the chance of motion from both mechanical instability and patient motion and effectively reduces the radiation dose required to obtain a 3D-DSA by half in those systems which require 2 rotations.

In the context of medical imaging, machine learning methods have been investigated since early 1990s[11-14], however the recent unprecedented performance of deep learning, has made major advances in solving very difficult problems in science, that were thought to be intractable when approached by other means [34, 35]. In addition to clever mathematical techniques and availability of large annotated datasets, many authors recognize that the massively parallel computing capabilities of GPUs have played a key role in the success of deep learning applications, providing accelerations of 40x – 250x compared to multicore and single core CPUs [10, 15]. For example, the training procedure of the network used in this study took approximately 23 hours. This training procedure could have taken 4-5 weeks if only a multicore CPU computation architecture was used, making this application unpractical. Fortunately, the training procedure is performed offline, it only needs to be done once and GPU computing is already widely available within the medical imaging community or accessible via cloud computing services such as the Google Cloud Platform (GCP) or the Amazon Web Service (AWS). Also, many standard open-source libraries used for deep learning applications, are highly optimized to be used in conjunction with GPUs. Once the parameters of the model have been learned, the process of analyzing new data that was not used for training the model (i.e. inference) can be further optimized for production. The method proposed in this study, uses a voxel-wise training and inference where the input of the network is a small image region of 41x41x5 voxels around the voxel of interest. This approach has multiple benefits, 1) inference can be parallelized, in other words, the classification of multiple voxels can be performed at the same time. Therefore, the time required to analyze a new case is directly proportional to the number of voxels to be classified (e.g. the entire head or a targeted ROI) and the number and generation of available GPUs. The throughput of the particular research implementation of the CNN model used in this study is approximately 2500 voxels/s/GPU. 2) This approach results in a large training dataset consisting of 150 million labeled voxels derived from 13790 axial images and 35 exams. In addition to a large training dataset, it is important to have a large testing cohort in order to assure a good model generalization that better reflects how this technique could be used in practice. Having a large testing cohort, also helps to determine whether the training dataset is large enough to achive a desire level of performance.

Although DLA images were successfully created for all validation and testing cases and were subject to quantitative and qualitative image analysis, this study still has some limitations: First, the use of a very specific image acquisition protocol and reconstruction with selective intra-arterial contrast media injection into the proximal internal carotid artery or vertebral artery may limit its clincal applications. Fine tuning of the model and clinical validations with prospective reader studies are required to further generalize these results to the vasculature of other organ systems, to complex or uncommon vascular abnormalities, as well as to 3D-DSA studies acquired using different image acquisition protocols and modalities (e.g. injection of IV contrast media, 4D-DSA, MDCT, etc.). This kind of prospective reader studies would also overcome the limitation of the current qualitative evaluation in our study by a single reader (Dr. Strother). Second, in this study a specific type of deep CNN with 30 layers, implemented with a ResNet [31, 32] architecture was used to demonstrate the DLA application. The selection of the well-known 30-layer ResNet is purely due to the fact that this architecture won an image contest among computer scientists (i.e. ILSVRC 2015 classification task), namely, it outperformed other types of networks such as AlexNet, VGGNet and GoogLeNet, for the task of natural images classification using the ImageNet data set. This type of network architecture has also outperformed other type of networks in medical imaging classification tasks with deeper models (i.e. increased number of layers) having improved classification accuracy [20, 22, 32, 36]. However, it remains unknown whether other architectures can be used for DLA and what would be the advanteages or disadvantages among all these networks. Furthernmore, additional optimization and fine tuning of the DLA model hyper-parameters (e.g. number of layers, number of hidden units per layer, learning rate, regularization schemes, etc.) is required for optimal online implementation and compatibility with clinical workflow. Third, even though metallic objects are automatically subtracted in the 3D-DSA images that were used to create the training dataset, small movements of metallic implants (e.g. an aneurysm clip or a coil mass) which occur during a cardiac cycle are, in the case of subtracted images, usually sufficient to create enough mis-registration artifacts to allow detection of an implant presence. This situation, in addition to the high x-ray attenuation and proximity to vasculature of metallic implants could result in their imitation in the final DLA images. The presence of high attenuating object (e.g. metal or Onyx) is also known to be an intrinsic limitation of mask-free angiography (vendors who provide a method to obtain 3D-DSAs without mask also offer the ability to perform a mask and fill acquisition in situations where metal objects are known to be present) and its clinical implications need to be addressed with an expert reader study.

**Conclusions**

A DLA method based in CNNs that generates 3D cerebral angiograms from a contrast enhanced C-arm cone beam CT without mask data acquisition was developed. Results indicate that the proposed method can successfully reduce mis-registration artifacts induced by inter-sweep patient motion and, by eliminating the need for a mask acquisition, reduces radiation dose in future clinical 3D angiography.

## Acknowledgments

The authors would like to thank Dr. John W. Garrett for grateful technical and editorial assistance.

## References

1. Go AS, Mozaffarian D, Roger VL, et al. Executive Summary: Heart Disease and Stroke Statistics—2013 Update. *A Report From the American Heart Association* 2013;127:143-152

2. Wiebers DO. Unruptured intracranial aneurysms: natural history, clinical outcome, and risks of surgical and endovascular treatment. *The Lancet* 2003;362:103-110

3. Mohr JP, Parides MK, Stapf C, et al. Medical management with or without interventional therapy for unruptured brain arteriovenous malformations (ARUBA): a multicentre, non-blinded, randomised trial. *The Lancet* 2014;383:614-621

4. Ogilvy CS, Stieg PE, Awad I, et al. AHA Scientific Statement: Recommendations for the management of intracranial arteriovenous malformations: a statement for healthcare professionals from a special writing group of the Stroke Council, American Stroke Association. *Stroke* 2001;32:1458-1471

5. Hacein-Bey L, Provenzale JM. Current imaging assessment and treatment of intracranial aneurysms. *AJR Am J Roentgenol* 2011;196:32-44

6. Anxionnat R, Bracard S, Ducrocq X, et al. Intracranial Aneurysms: Clinical Value of 3D Digital Subtraction Angiography in the Therapeutic Decision and Endovascular Treatment. *Radiology* 2001;218:799-808

7. Gandhi D, Chen J, Pearl M, et al. Intracranial dural arteriovenous fistulas: classification, imaging findings, and treatment. *AJNR Am J Neuroradiol* 2012;33:1007-1013

8. Fahrig R, Fox AJ, Lownie S, et al. Use of a C-arm system to generate true three-dimensional computed rotational angiograms: preliminary in vitro and in vivo results. *American Journal of Neuroradiology* 1997;18:1507-1514

9. Strobel N, Meissner O, Boese J, et al. 3D Imaging with Flat-Detector C-Arm Systems. In: Reiser MF, Becker CR, Nikolaou K, et al., eds. *Multislice CT*. Berlin, Heidelberg: Springer Berlin Heidelberg; 2009:33-51

10. Erickson BJ, Korfiatis P, Akkus Z, et al. Machine Learning for Medical Imaging. *RadioGraphics* 2017;37:505-515

11. Chan H-P, B. LS-C, Berkman S, et al. Computer-aided detection of mammographic microcalcifications: Pattern recognition with an artificial neural network. *Medical Physics* 1995;22:1555--1567

12. Wu Y, Kunio D, L. GM, et al. Computerized detection of clustered microcalcifications in digital mammograms: Applications of artificial neural networks. *Medical Physics* 1992;19:555--560

13.	Zhang W, Kunio D, L. GM, et al. Computerized detection of clustered microcalcifications in digital mammograms using a shift-invariant artificial neural network. *Medical Physics* 1994;21:517--524

14.	Wang S, Summers RM. Machine learning and radiology. *Medical Image Analysis* 2012;16:933-951

15.	LeCun Y, Bengio Y, Hinton G. Deep learning. *Nature* 2015;521:436-444

16.	Gulshan V, Peng L, Coram M, et al. Development and validation of a deep learning algorithm for detection of diabetic retinopathy in retinal fundus photographs. *JAMA* 2016;316:2402-2410

17.	Huynh BQ, Li H, Giger ML. Digital mammographic tumor classification using transfer learning from deep convolutional neural networks. *Journal of Medical Imaging* 2016;3:034501

18.	Antropova N, Q. HB, L. GM. A deep feature fusion methodology for breast cancer diagnosis demonstrated on three imaging modality datasets. *Medical Physics* 2017;[Epub ahead of print]

19.	Akkus Z, Galimzianova A, Hoogi A, et al. Deep Learning for Brain MRI Segmentation: State of the Art and Future Directions. *Journal of Digital Imaging* 2017;30:449-459

20.	Korfiatis P, Kline TL, Lachance DH, et al. Residual Deep Convolutional Neural Network Predicts MGMT Methylation Status. *Journal of Digital Imaging* 2017

21.	Nakao T, Hanaoka S, Nomura Y, et al. Deep neural network-based computer-assisted detection of cerebral aneurysms in MR angiography. *Journal of Magnetic Resonance Imaging* 2017 [Epub ahead of print]

22. Wang X, Peng Y, Lu L, et al. ChestX-ray8: Hospital-Scale Chest X-Ray Database and Benchmarks on Weakly-Supervised Classification and Localization of Common Thorax Diseases. *2017 IEEE Conference on Computer Vision and Pattern Recognition (CVPR)*; 2017

23. Lakhani P, Sundaram B. Deep Learning at Chest Radiography: Automated Classification of Pulmonary Tuberculosis by Using Convolutional Neural Networks. *Radiology* 2017;284:574-582

24. Cybenko G. Approximation by superpositions of a sigmoidal function. *Mathematics of Control, Signals and Systems* 1989;2:303-314

25. Hornik K. Approximation capabilities of multilayer feedforward networks. *Neural Networks* 1991;4:251-257

26. Struffert T, Hauer M, Banckwitz R, et al. Effective dose to patient measurements in flat-detector and multislice computed tomography: a comparison of applications in neuroradiology. *European Radiology* 2014;24:1257-1265

27. Lang S, Gölitz P, Struffert T, et al. 4D DSA for Dynamic Visualization of Cerebral Vasculature: A Single-Center Experience in 26 Cases. *American Journal of Neuroradiology* 2017;38:1169

28. Shapiro LG. Connected Component Labeling and Adjacency Graph Construction. *Machine Intelligence and Pattern Recognition* 1996;19:1-30

29. He H, Garcia EA. Learning from Imbalanced Data. *IEEE Transactions on Knowledge and Data Engineering* 2009;21:1263-1284

30. LeCun Y, Boser BE, Denker JS, et al. Handwritten Digit Recognition with a Back-Propagation Network. *Advances in Neural Information Processing Systems*; 1990:396--404

31. He K, Zhang X, Ren S, et al. Identity Mappings in Deep Residual Networks. In: Leibe B, Matas J, Sebe N, et al., eds. *Computer Vision – ECCV 2016: 14th European Conference, Amsterdam, The Netherlands, October 11–14, 2016, Proceedings, Part IV*. Cham: Springer International Publishing; 2016:630-645

32. He K, Zhang X, Ren S, et al. Deep Residual Learning for Image Recognition. *2016 IEEE Conference on Computer Vision and Pattern Recognition (CVPR)*; 2016:770-778

33. He K, Zhang X, Ren S, et al. Delving Deep into Rectifiers: Surpassing Human-Level Performance on ImageNet Classification. *2015 IEEE International Conference on Computer Vision (ICCV)*; 2015:1026-1034

34. Silver D, Huang A, Maddison CJ, et al. Mastering the game of Go with deep neural networks and tree search. *Nature* 2016;529:484-503

35. Silver D, Schrittwieser J, Simonyan K, et al. Mastering the game of Go without human knowledge. *Nature* 2017;550:354

36. Shin HC, Roth HR, Gao M, et al. Deep Convolutional Neural Networks for Computer-Aided Detection: CNN Architectures, Dataset Characteristics and Transfer Learning. *IEEE Transactions on Medical Imaging* 2016;35:1285-1298

**Table 1.** Summary of performance metrics for vascular classification in the training, validation and testing datasets.

| Dataset | Sensitivity (recall) $\frac{TP}{TP+FN}$ | | PPV (precision) $\frac{TP}{TP+FP}$ | | DSC (F1-score) $\frac{2TP}{2TP+FP+FN}$ | | Accuracy $\frac{TP+TN}{TP+TN+FP+FN}$ | |
|---|---|---|---|---|---|---|---|---|
| | Mean | 95% CI | Mean | 95% CI | Mean | 95% CI | Mean | 95% CI |
| Validation (n=8) | 97.8% | [96.9, 98.7]% | 97.2% | [96.4, 98.1]% | 97.5% | [97.0, 98.0]% | 98.4% | [98.0, 98.7]% |
| Testing (n=62) | 98.5% | [97.6, 99.3]% | 97.6% | [96.5, 98.6]% | 98.0% | [97.4, 98.7]% | 98.7% | [98.3, 99.1]% |

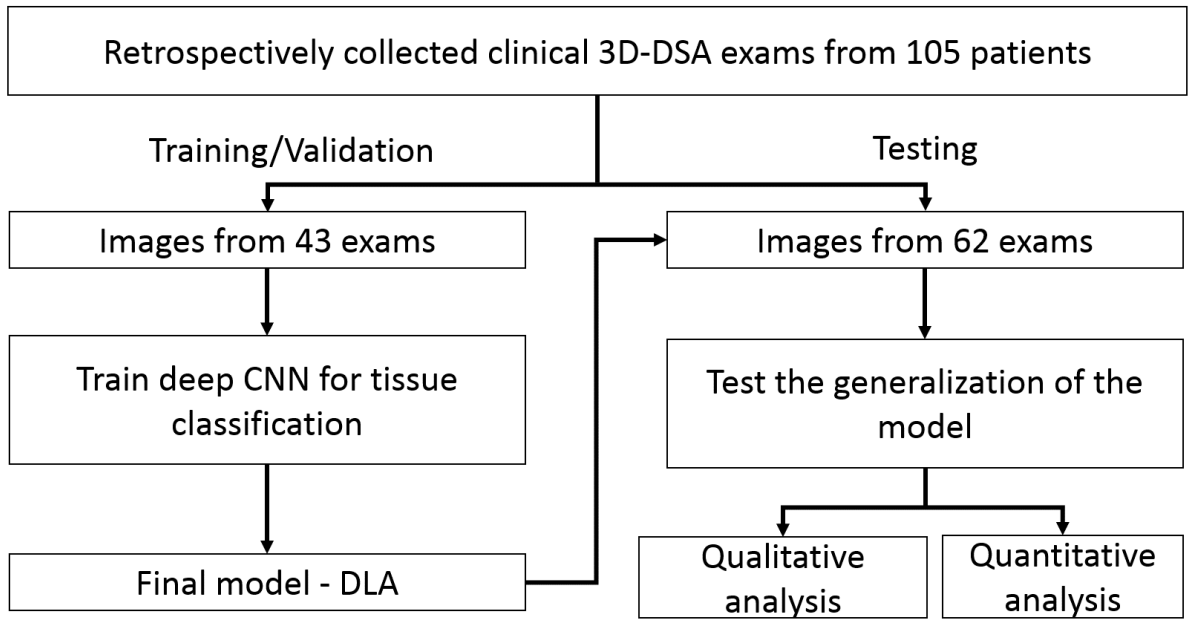

**Figure 1.** Overall study schema.

**Figure 2.** Neural Network Architecture.

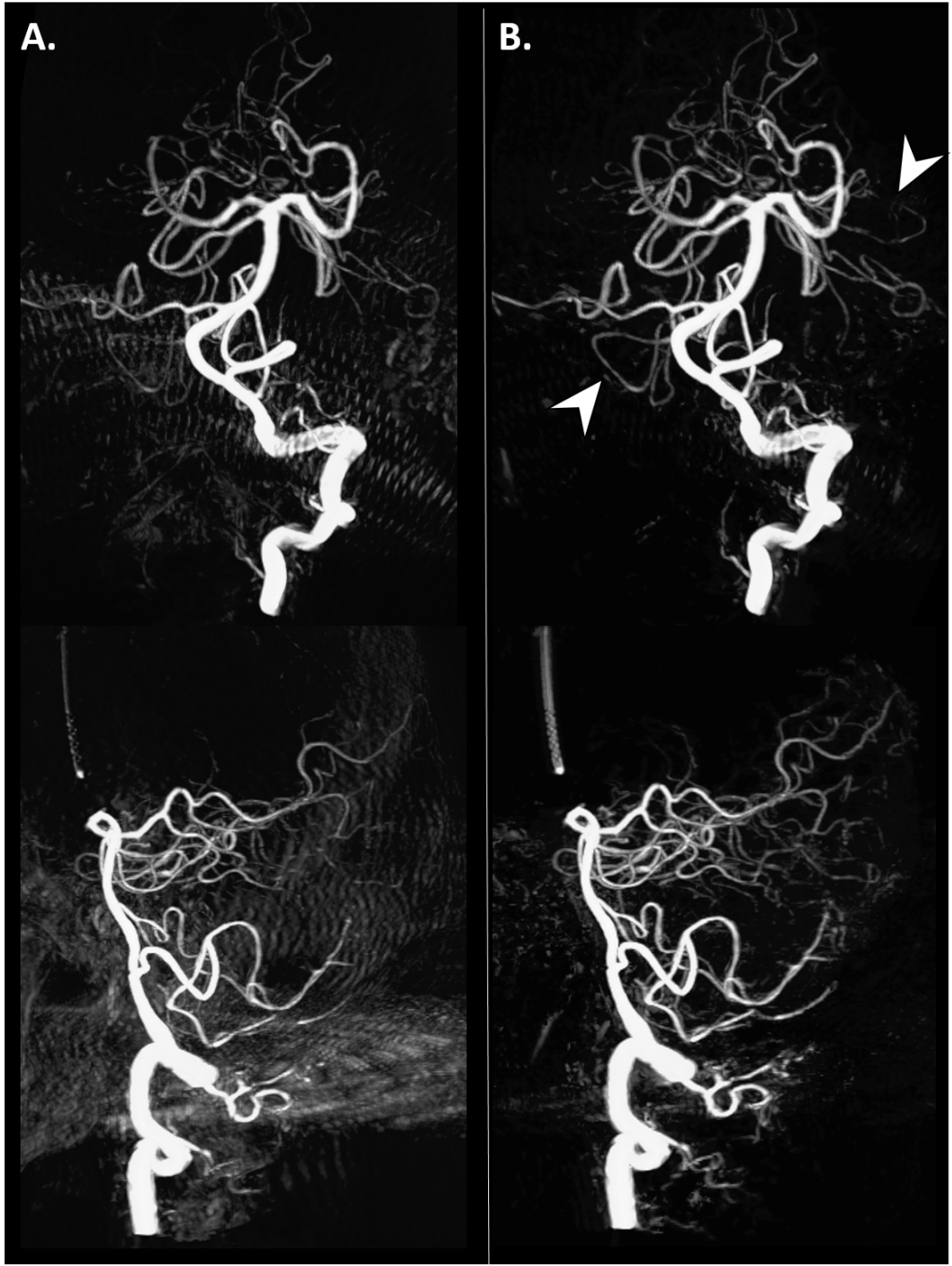

**Figure 3.** Comparison of anterior and lateral views of MIP images derived from A) 3D-DSA and B) 3D-DLA datasets of a patient evaluated for posterior circulation. Residual bone artifacts induced by inter-sweep patient motion are greatly reduced in 3D-DLA, improving the conspicuity of small vessels as pointed by the white arrows.

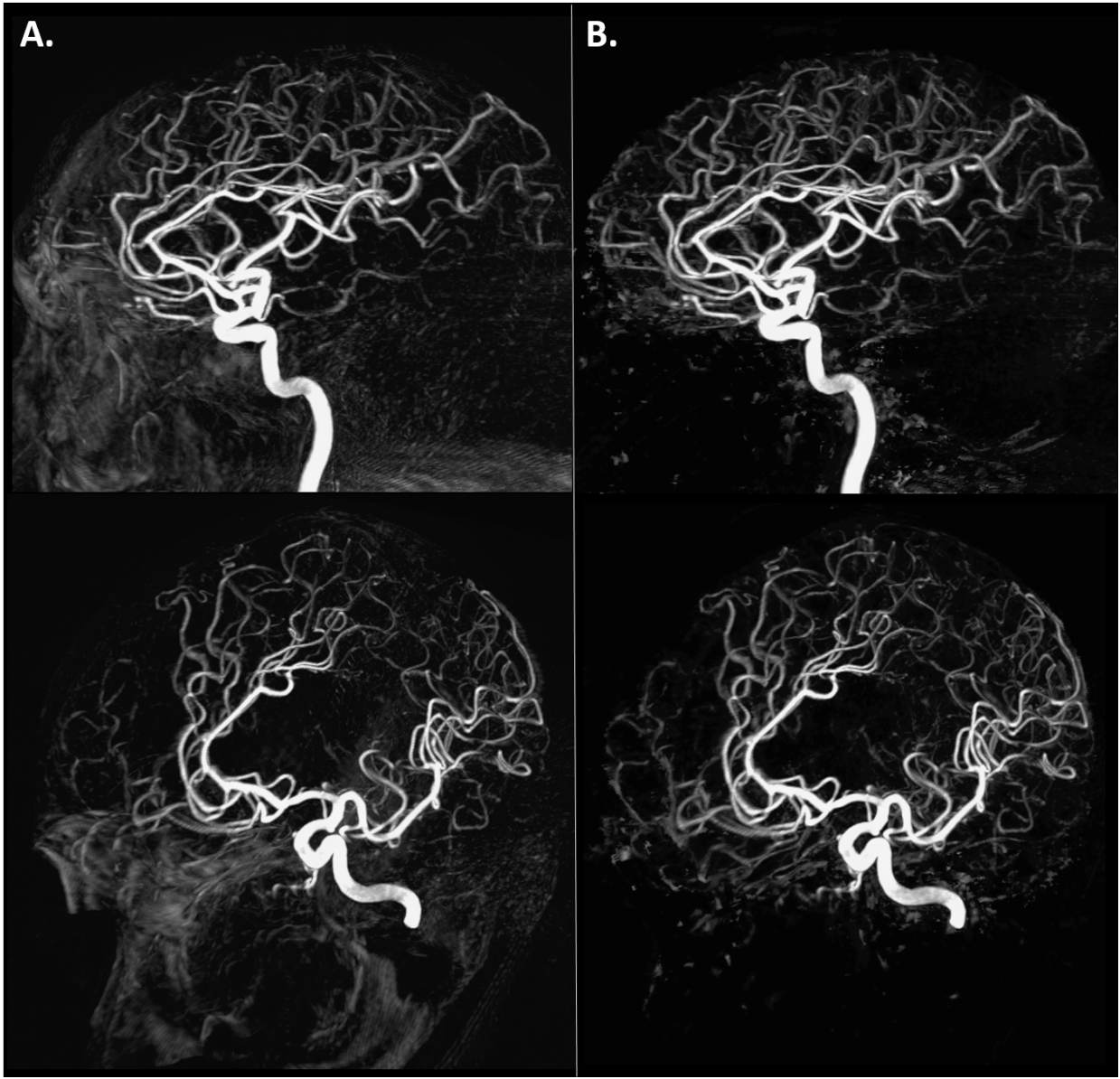

**Figure 4.** Comparison of lateral and oblique MIP images derived from A) 3D-DSA and B) 3D-DLA datasets of a patient evaluated for anterior circulation. Results show reduced residual bone artifacts for 3D-DLA images, in particular for the anterior aspect of the skull and the temporal bone.

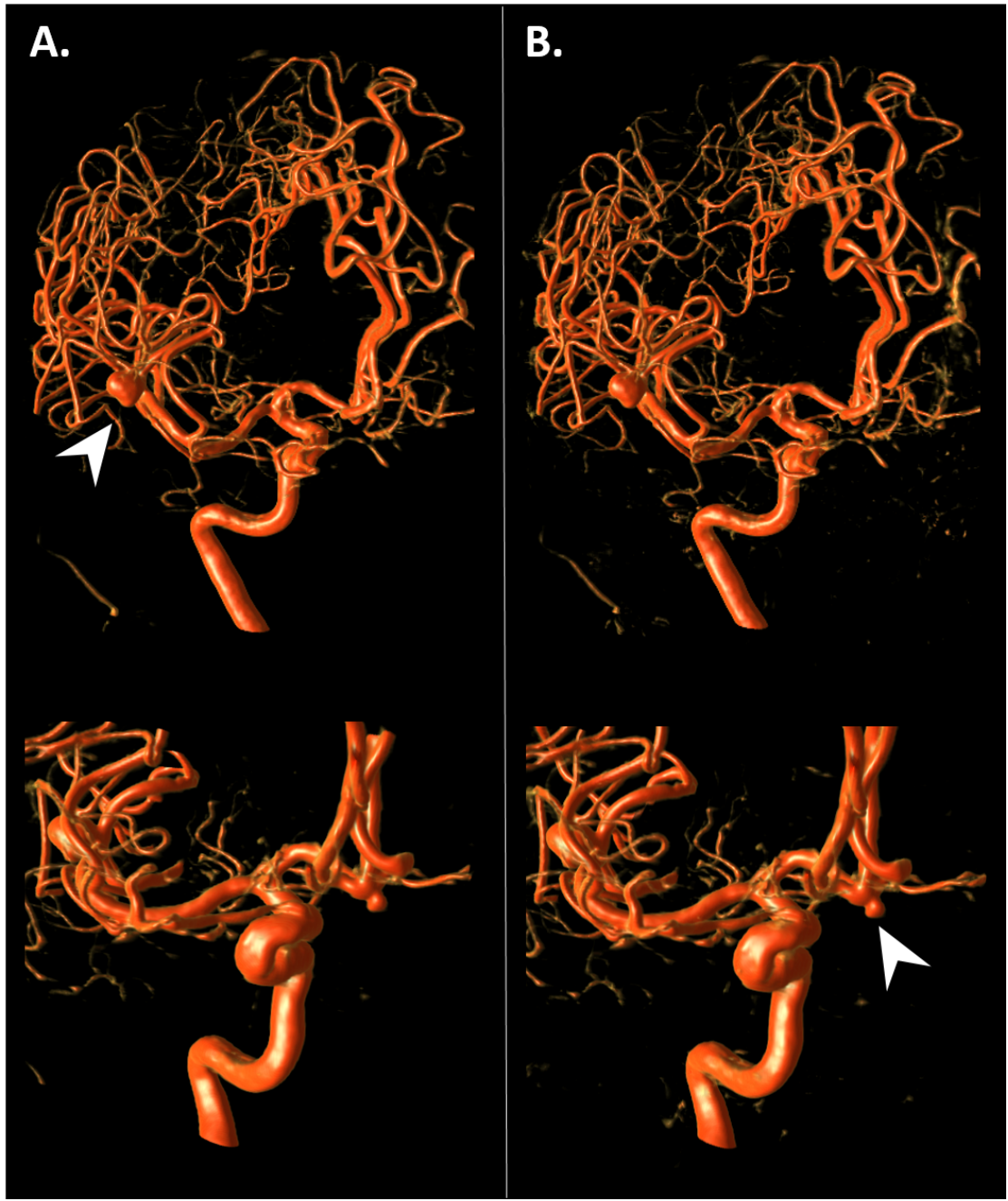

**Figure 5.** Comparison of volume rendering images for both the A) clinical 3D-DSA and B) the 3D-DLA of a patient with a small aneurysm in the anterior communicating artery and a large aneurysm in the MCA bifurcation as pointed by the arrows.